\documentclass[reprint,amssymb, amsmath, aps, superscriptaddress, showpacs, footinbib, prl]{revtex4-1}
\pdfoutput=1
\usepackage{graphicx}
\usepackage{color}

\newcommand{\eff}{\text{eff}}
\newcommand{\AFM}{\text{AFM}}

\newcommand{\mg}{\text{mag}}

\newcommand{\kv}{\mathbf k}
\newcommand{\Sv}{\mathbf S}
\newcommand{\rv}{\mathbf r}

\begin{document}

\title{Collinear order in a frustrated three-dimensional spin-$\frac12$ antiferromagnet Li$_2$CuW$_2$O$_8$}

\author{K. M. Ranjith}

\author{R. Nath}
\email{rnath@iisertvm.ac.in}
\affiliation{School of Physics, Indian Institute of Science Education and Research Thiruvananthapuram-695016, India}

\author{M. Skoulatos}
\affiliation{Laboratory for Neutron Scattering and Imaging, Paul Scherrer Insitut, CH-5232 Villigen PSI, Switzerland}
\affiliation{Department of Physics, Technical University Munich, 85748 Garching, Germany}

\author{L. Keller}
\affiliation{Laboratory for Neutron Scattering and Imaging, Paul Scherrer Insitut, CH-5232 Villigen PSI, Switzerland}

\author{D. Kasinathan}
\affiliation{Max Planck Institute for Chemical Physics of Solids, 01187 Dresden, Germany}

\author{Y.~Skourski}
\affiliation{Dresden High Magnetic Field Laboratory, Helmholtz-Zentrum Dresden-Rossendorf, 01314 Dresden, Germany}

\author{A. A. Tsirlin}
\email{altsirlin@gmail.com}
\affiliation{Max Planck Institute for Chemical Physics of Solids, 01187 Dresden, Germany}
\affiliation{National Institute of Chemical Physics and Biophysics, 12618 Tallinn, Estonia}
\affiliation{Experimental Physics VI, Center for Electronic Correlations and Magnetism, Institute of Physics, University of Augsburg, 86135 Augsburg, Germany}


\begin{abstract}
Magnetic frustration in three dimensions (3D) manifests itself in the spin-$\frac12$ insulator Li$_2$CuW$_2$O$_8$. Density-functional band-structure calculations reveal a peculiar spin lattice built of triangular planes with frustrated interplane couplings. The saturation field of 29\,T contrasts with the susceptibility maximum at 8.5\,K and a relatively low N\'eel temperature $T_N\simeq 3.9$\,K. Magnetic order below $T_N$ is collinear with the propagation vector $(0,\frac12,0)$ and an ordered moment of 0.65(4)\,$\mu_B$ according to neutron diffraction data. This reduced ordered moment together with the low maximum of the magnetic specific heat ($C^{\max}/R\simeq 0.35$) pinpoint strong magnetic frustration in 3D. Collinear magnetic order suggests that quantum fluctuations play crucial role in this system, where a non-collinear spiral state would be stabilized classically.
\end{abstract}

\maketitle

\textit{Introduction}. Magnetic frustration, the competition of exchange couplings between localized spins, has broad implications for ground states, excitation spectra, and low-temperature properties. Prominent manifestations of the frustration include peculiar behaviors of spin ice~\cite{gingras2014}, the formation of quantum spin liquids~\cite{balents2010}, and non-collinear magnetic structures that give rise to spin chirality and strong magnetoelectric coupling~\cite{mostovoy2007}. The properties of frustrated magnets change drastically depending on the dimensionality of the spin lattice. Rigorous mapping between theory and experiment requires that both magnetic models and real materials are relatively simple. In this context, the case of isotropic (Heisenberg) exchange on a three-dimensional (3D) lattice of quantum spin-$\frac12$ ions is perhaps least studied, given the lack of model materials and the complexity of the numerical treatment of purely quantum spins in 3D. On the materials side, 3D spin-$\frac12$ frustrated magnetism has been proposed for the hyperkagome lattice~\cite{lawler2008,*zhou2008,*bergholtz2010,dally2014} in Na$_4$Ir$_3$O$_8$~\cite{okamoto2007}, but the involvement of the $5d$ Ir$^{4+}$ ion with its inherently strong spin-orbit coupling may lead to deviations from the isotropic Heisenberg regime~\cite{chen2008,norman2010,*micklitz2010}. A similar hyperkagome-like exchange network was recently reported in a $3d$-based material PbCuTe$_2$O$_6$~\cite{koteswararao2014}.

Here, we report a long-sought 3D frustrated magnet built of \mbox{spin-$\frac12$} Cu$^{2+}$ ($3d$) ions that feature nearly isotropic Heisenberg exchange with interactions frustrated along all three crystallographic directions. This  frustration manifests itself in thermodynamic properties, whereas neutron diffraction reveals a collinear magnetic ground state, thus granting us valuable insight into the properties of a 3D frustrated spin-$\frac12$ Heisenberg antiferromagnet. We conclude that the collinear state is stabilized in this system, even though a non-collinear state should have lower energy on the classical level. 

\begin{figure}
\includegraphics{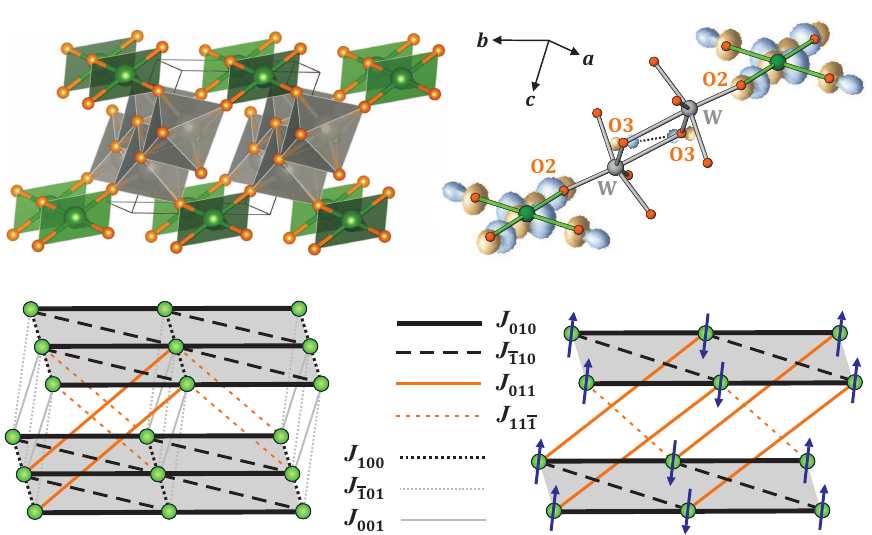}
\caption{\label{fig:structure}
(Color online). Top: crystal structure of Li$_2$CuW$_2$O$_8$ with Li atoms omitted for clarity (left), and the superexchange pathway for one of the long-range coupling, $J_{011}$, revealed by Wannier functions (right). Bottom: spin lattice of Li$_2$CuW$_2$O$_8$ (left) and the experimental magnetic ground state (right). Triangular planes are shaded. All couplings are antiferromagnetic. In the right panel, only those couplings that stabilize the experimental magnetic structure are shown.
}
\end{figure}
Our work is focused on Li$_2$CuW$_2$O$_8$ featuring magnetic spin-$\frac12$ Cu$^{2+}$ ions and non-magnetic W$^{6+}$. The crystal structure of Li$_2$CuW$_2$O$_8$~\cite{alvarez2001} depicted in the top part of Fig.~\ref{fig:structure} entails planar CuO$_4$ plaquette units (green) that are held together by WO$_6$ octahedra (gray). Despite the triclinic symmetry of the crystal structure (space group $P\bar 1$), we expect a relatively simple topology of magnetic interactions, because the unit cell contains only one Cu atom located at the inversion center. Moreover, the presence of inversion centers in the middle of each Cu--Cu bond implies that the Dzyaloshinsky-Moriya (DM) anisotropy, which is the leading anisotropy term in Cu$^{2+}$ compounds~\cite{shekhtman1992,*shekhtman1993}, \emph{vanishes by symmetry} in contrast to PbCuTe$_2$O$_6$~\cite{koteswararao2014}, where a twisted arrangement of the CuO$_4$ plaquettes favors the antisymmetric DM exchange. This ensures reliable mapping between experimental results and theory, which is typically developed for the Heisenberg Hamiltonians with isotropic exchange.

\textit{Microscopic magnetic model.} Individual magnetic couplings were quantified by DFT calculations performed using the \texttt{FPLO} code~\cite{fplo} within the local-density approximation (LDA) for the exchange-correlation potential~\cite{pw92}. Antiferromagnetic (AFM) contributions to the exchange integrals are obtained as $J_i^{\AFM}=4t_i^2/U_{\eff}$, where $t_i$ are electron hoppings extracted from the LDA band structure, and $U_{\eff}=4.5$\,eV is an effective on-site Coulomb repulsion~\cite{lebernegg2013,*nath2014}. Alternatively, we evaluated total exchange integrals $J_i$ as energy differences between collinear spin states in LSDA+$U$, where a mean-field correction for on-site correlation effects is added to the LDA functional. The local correlation parameters are $U_d=7.5$\,eV and $J_d=1$\,eV for the Coulomb repulsion and Hund's exchange, respectively~\cite{[{We used the around-mean-field double-counting correction, similar to: }][{}]lebernegg2011,*lebernegg2014}. The change in the $U_d$ parameter and/or double-counting correction scheme had no qualitative effect on the model, although quantitative agreement with the experiment became less satisfactory.

\begin{figure}
\includegraphics{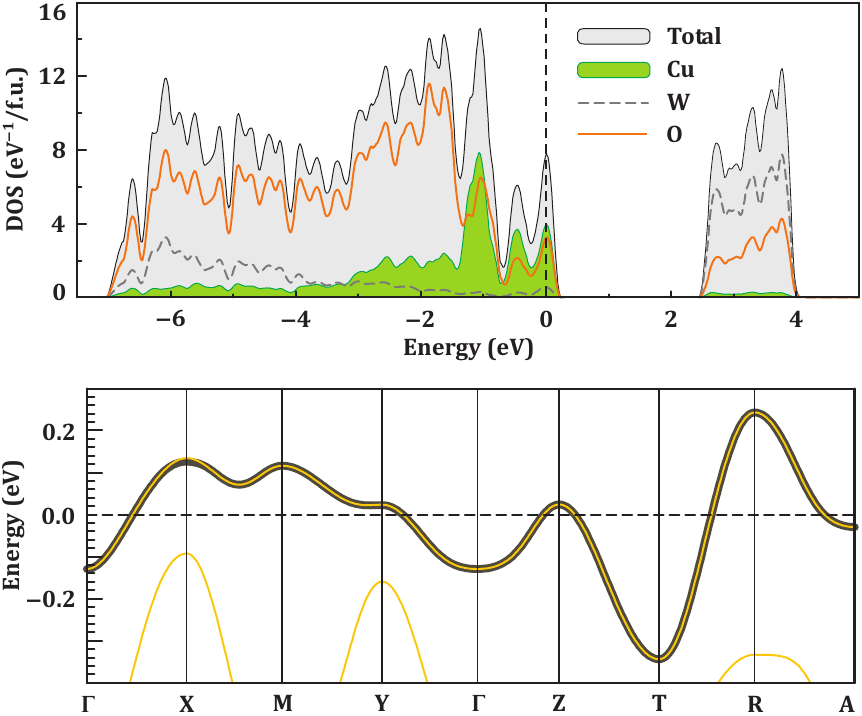}
\caption{\label{fig:dos}
(Color online). Top: LDA density of states for Li$_2$CuW$_2$O$_8$. Bottom: LDA bands for Li$_2$CuW$_2$O$_8$ (thin light lines) and the fit of the Cu $d_{x^2-y^2}$ band with the tight-binding model (thick dark line). The Fermi level is at zero energy. Note that the band structure is metallic because strong correlation effects in the Cu $3d$ shell are nearly absent in LDA.
}
\end{figure}

LDA density of states and band structure (Fig.~\ref{fig:dos}) evidence metallic behavior because strong correlation effects in the Cu $3d$ shell are largely underestimated in LDA. When LSDA+$U$ is used, the band gap of about 3.0\,eV and magnetic moments of about 0.86\,$\mu_B$ on Cu atoms are obtained. Experimentally, pale-yellow-colored Li$_2$CuW$_2$O$_8$ is clearly insulating. However, no quantitative information on its electronic structure is available.

\begin{table}
\caption{\label{tab:exchanges}
\textit{Ab initio} estimates of the exchange couplings in Li$_2$CuW$_2$O$_8$: interatomic distances $d_{\text{Cu--Cu}}$ (in\,\r A), electron hoppings $t_i$ (in\,meV), AFM contributions to the exchange $J_i^{\AFM}$ (in\,K) obtained as $4t_i^2/U_{\eff}$, and total exchange integrals $J_i$ (in\,K) from LSDA+$U$ calculations.
}
\begin{ruledtabular}
\begin{tabular}{ccrrr}
                  & $d_{\text{Cu--Cu}}$ & $t_i$ & $J_i^{\AFM}$ & $J_i$ \\
  $J_{100}$       &  4.967 & $-21$ & 5  & 2.9  \\
  $J_{010}$       &  5.497 & $-38$ & 15 & 10.6 \\
  $J_{\bar 110}$  &  5.719 &   20  & 4  & 4.0  \\
  $J_{001}$       &  5.888 &   16  & 3  & 0.3  \\
  $J_{\bar 101}$  &  7.433 & $-16$ & 3  & 2.2  \\  
  $J_{011}$       &  9.288 &   28  & 8  & 5.1  \\
  $J_{11\bar 1}$  &  9.288 & $-27$ & 8  & 1.1  \\
\end{tabular}
\end{ruledtabular}
\end{table}
\begin{table}
\caption{\label{tab:exchanges-suppl}
Exchange couplings in Li$_2$CuW$_2$O$_8$ (in\,K) obtained with different values of the $U_d$ parameter in LSDA+$U$. Ensuing macroscopic parameters, the Curie-Weiss temperature $\theta$ (in\,K) and saturation field $H_s$ (in\,T), are listed as well. The bottom line is the energy difference between the experimental collinear spin configuration and the incommensurate classical ground state, $E_{\rm noncol}-E_{\rm col}$ (in\,K). The variation of $U_d$ does not change the microscopic scenario qualitatively, although a good quantitative agreement with the experiment is achieved at $U_d=7.5$\,eV only.
}
\begin{ruledtabular}
\begin{tabular}{cccc}
                & $U_d=6.5$\,eV & $U_d=7.5$\,eV & $U_d=8.5$\,eV \\
 $J_{100}$      &   3.6         &     2.9       &      2.7      \\
 $J_{010}$      &   12.7        &    10.6       &      9.4      \\
 $J_{\bar 110}$ &   4.9         &    4.0        &      3.6      \\
 $J_{001}$      &   0.6         &    0.3        &      0.3      \\
 $J_{\bar 101}$ &   2.9         &    2.2        &      2.0      \\
 $J_{011}$      &   6.2         &    5.1        &      4.4      \\
 $J_{11\bar 1}$ &   1.3         &    1.1        &      0.8      \\\hline
 $\theta$       &  16.1         &   13.1        &     11.6      \\
 $H_s$          &  34.5         &   28.6        &     25.0      \\
 $E_{\rm noncol}-E_{\rm col}$ & $-1.0$ & $-0.5$ &   $-0.7$      \\
\end{tabular}
\end{ruledtabular}
\end{table}
Exchange couplings in Li$_2$CuW$_2$O$_8$ are listed in Table~\ref{tab:exchanges}. Taking advantage of the single Cu$^{2+}$ ion in the unit cell, we label all $J$'s according to their relevant crystallographic directions, see Fig.~\ref{fig:structure}. Triclinic symmetry of the crystal structure implies that only $\mathbf r$ and $-\mathbf r$ are equivalent, whereas none of the face or body diagonals of the unit cell are related by symmetry. The effect of the on-site Coulomb repulsion parameter $U_d$ on the magnetic parameters can be seen from Table~\ref{tab:exchanges-suppl}.

\begin{figure}
\includegraphics{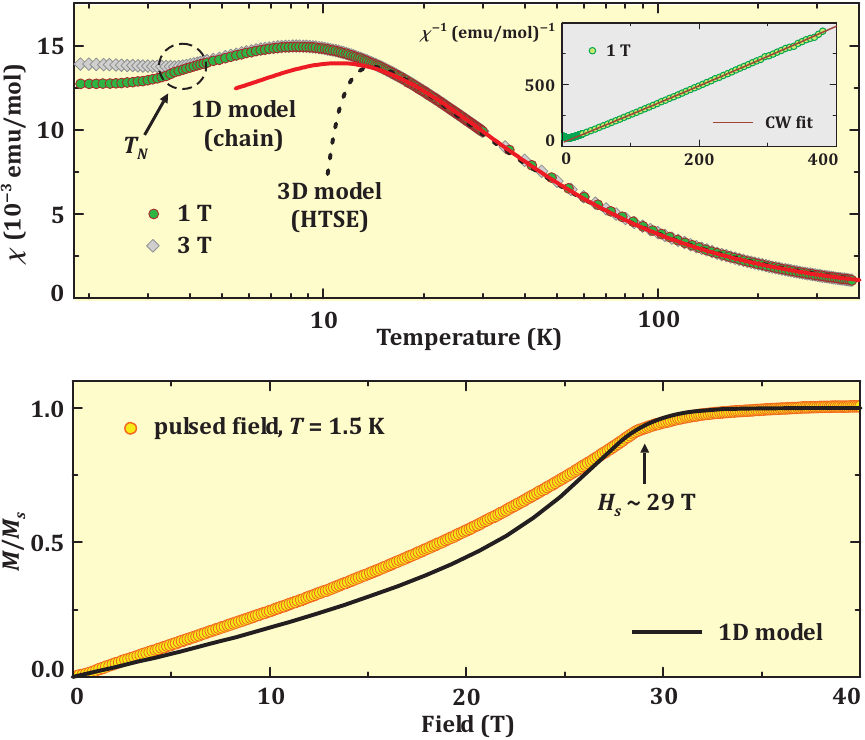}
\caption{\label{fig:chi}
(Color online). Top: magnetic susceptibility ($\chi$) of Li$_2$CuW$_2$O$_8$ as a function of temperature. The bifurcation of the 1\,T and 3\,T data around 4\,K indicates the magnetic ordering at $T_N\simeq 3.9$\,K. The fit of the 1D model (solid line) fails to reproduce the position of the susceptibility maximum. The fit with the 8th-order HTSE down to 13\,K is shown by the dashed line. The inset shows the Curie-Weiss fit above 80\,K. Bottom: magnetization curve of Li$_2$CuW$_2$O$_8$ measured at 1.5\,K in pulsed magnetic fields. The solid line is the fit of the 1D model.
}
\end{figure}
Sizable exchange couplings on the order of several K are found for Cu--Cu distances up to 10\,\r A. The leading coupling is along the $b$ direction ($J_{010}$) through a single WO$_6$ octahedron. However, several couplings mediated by two contiguous WO$_6$ octahedra ($J_{001}$, $J_{\bar 101}$, $J_{011}$, $J_{11\bar 1}$) are only 2-3 times weaker than $J_{010}$. These long-range couplings originate from the contributions of second-neighbor oxygen atoms to Wannier functions centered on Cu sites. For example, the sizable contributions of the $p$-orbitals of O3 are responsible for the coupling $J_{011}$ between those Cu atoms that are more than 9\,\r A apart (Fig.~\ref{fig:structure}, top right).

The couplings $J_{010}$, $J_{100}$, and $J_{\bar 110}$ form a frustrated triangular lattice in the $ab$ plane (Fig.~\ref{fig:structure}, bottom). $J_{011}$ is the leading interaction along $c$, but several other interactions are present as well, and two of them, $J_{001}$ and $J_{\bar 101}$, are not compatible with $J_{011}$. They form triangular loops and generate additional frustration. Reduction to the purely one-dimensional (1D) model by retaining only the leading coupling $J_{010}$ could be envisaged, but the 1D behavior is not observed experimentally as we show below. On the other hand, Li$_2$CuW$_2$O$_8$ can not be viewed as a pure triangular-lattice system because the leading interplane coupling $J_{011}$ is as strong as the couplings $J_{100}$ and $J_{\bar 110}$ within the plane.

\textit{Experimental techniques.} In order to probe the magnetism of Li$_2$CuW$_2$O$_8$ experimentally, we prepared polycrystalline samples by firing stoichiometric mixtures of Li$_2$CO$_3$, CuO, and WO$_3$ at 650\,$^{\circ}$C for 24\,h and subsequently at 750\,$^{\circ}$C for 48\,h. Sample quality was checked by powder x-ray diffraction (Empyrean diffractometer from PANalytical, CuK$_{\alpha}$ radiation), and no impurity phases were found. Thermodynamic properties were measured as a function of temperature ($T$) and field ($H$) using the Quantum Design MPMS SQUID and PPMS devices. High-field magnetization measurements were performed at the Dresden High Magnetic Field Laboratory in pulsed fields. 

\textit{Magnetization.} Magnetic susceptibility ($\chi$) of Li$_2$CuW$_2$O$_8$ reveals a broad maximum around 8.5\,K (Fig.~\ref{fig:chi}, top, inset). A weak kink around 4\,K and a bifurcation of the susceptibility curves measured in different fields indicate the long-range AFM ordering around $T_N\simeq 3.9$\,K. This behavior resembles quasi-1D magnets, where the broad maximum of the susceptibility above $T_N$ is due to the short-range order in 1D. However, attempts to fit the experimental susceptibility curve with the standard expression for the uniform spin-$\frac12$ chain~\cite{johnston2000} ultimately failed, and even the position of the susceptibility maximum could not be adequately reproduced (Fig.~\ref{fig:chi}, top). This indicates  magnetic frustration by interchain couplings that impede short-range order, thus shifting the susceptibility maximum to lower temperatures~\cite{[{For example: }][]johnston2011}.

Magnetization ($M$) of Li$_2$CuW$_2$O$_8$ saturates around $H_s\simeq 29$\,T (Fig.~\ref{fig:chi}, bottom). Taking $J_{\rm 1D}\simeq 17$\,K from the susceptibility fit, we are able to reproduce the saturation field but not the magnetization isotherm itself~\cite{[{Magnetization curve and specific heat of the uniform spin-$\frac12$ chain were obtained from quantum Monte-Carlo simulations performed in the ALPS package: }][{}]alps}. Its curvature is much smaller than expected for the 1D model. The nearly linear magnetization curve is consistent with the proposed 3D nature of Li$_2$CuW$_2$O$_8$, since the curvature of $M(H)$ is reduced when the dimensionality of the spin lattice is increased~\cite{[{See, for example, Fig. 3 in: }][]goddard2012}.

Both $\chi(T)$ and $M(H)$ data show that the purely 1D model does not account for the physics of Li$_2$CuW$_2$O$_8$ even well above $T_N$, and this compound can not be considered as a quasi-1D magnet. Unfortunately, thermodynamics of the full 3D quantum spin model of Li$_2$CuW$_2$O$_8$ is beyond the reach of present-day numerical techniques because of the strong frustration. However, a comparison between DFT and experiment is possible on the mean-field level.
 
At high temperatures, $1/\chi(T)$ is linear and follows the Curie-Weiss law with an effective moment of 1.88\,$\mu_B$ and Curie-Weiss temperature $\theta\simeq 12$\,K (Fig.~\ref{fig:chi}, top). The $\theta$ value is a sum of individual exchange couplings. For a spin-$\frac12$ system, $\theta=\frac14\sum_iz_iJ_i$, where $z_i=2$ is the number of couplings per site. Using $J_i$'s from Table~\ref{tab:exchanges}, we arrive at $\theta\simeq 13$\,K in excellent agreement with the experiment. The saturation field is proportional to the couplings on the bonds, where spins have to be flipped in order to transform the AFM zero-field ground state into a fully polarized (ferromagnetic) state. Considering the experimental magnetic structure of Li$_2$CuW$_2$O$_8$ (Fig.~\ref{fig:structure}, bottom right), we expect 
\begin{equation}
H_s=2k_B/(g\mu_B)\times (J_{010}+J_{\bar 110}+J_{011}+J_{11\bar 1})\simeq 29\text{\,T},
\end{equation}
which is in excellent agreement with the high-field magnetization experiment. Here, we used an effective $g$-factor $g=2.17$ extracted from the paramagnetic effective moment $\mu_{\eff}=g\sqrt{S(S+1)}\simeq 1.88$\,$\mu_B$.

Additionally, we were able to fit the magnetic susceptibility above 13\,K using the 8th-order high-temperature series expansion (HTSE)~\cite{htse,*htsecode} with fixed ratios of individual exchange couplings and the leading exchange coupling $J_{010}$ as the variable parameter yielding $J_{010}\simeq 10.6$\,K in excellent agreement with the DFT results in Table~\ref{tab:exchanges}. Below 13\,K, the HTSE diverges.
\begin{figure}
\includegraphics{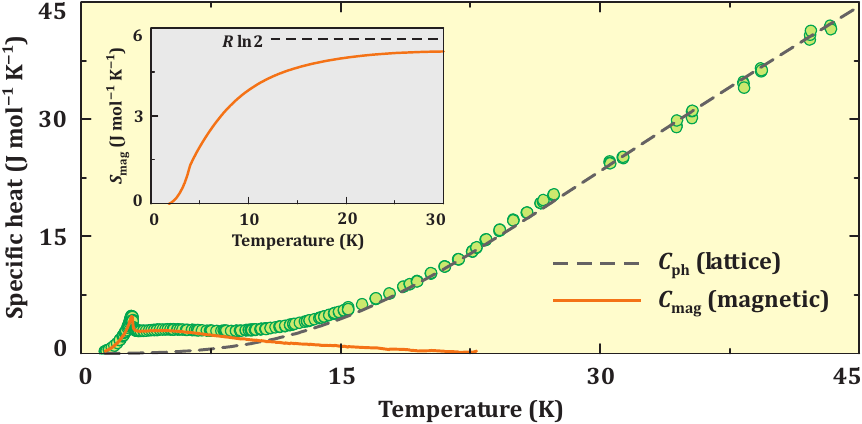}
\caption{\label{fig:subtraction} Temperature dependence of heat capacity measured in zero field for Li$_{2}$CuW$_{2}$O$_8$. The spheres are the raw data. The dashed line is the phonon contribution $C_{\rm ph}$ as found from the fit to Eq.~\eqref{Debye}, and the solid line denotes the magnetic contribution $C_{\rm mag}$. The inset shows the magnetic entropy $S_{\rm mag}$ as a function of $T$. The dashed horizontal line is the value $S_{\rm mag}=R\ln 2$ expected for Cu$^{2+}$ spins.}
\end{figure}

\textit{Specific heat}. Heat capacity ($C_p$) was measured on a small pressed pellet using the relaxation technique. In order to estimate the phonon contribution to the specific heat $C_{\rm ph}(T)$, the $C_{\rm p}(T)$ data above 40\,K were fitted by a sum of Debye contributions:
\begin{equation}
\label{Debye}
C_{\rm ph}(T) = 9R\displaystyle\sum\limits_{n=1}^{3} c_n \left(\frac{T}{\theta_{Dn}}\right)^3 \int_0^{\frac{\theta_{Dn}}{T}} \frac{x^4e^x}{(e^x-1)^2} dx,
\end{equation}
where $\theta_{Dn}$ are characteristic Debye temperatures and $c_n$ are integer coefficients indicating the contributions of different atoms (or groups of atoms) to $C_{\rm p}(T)$. A similar procedure has been adopted in several recent studies of quantum magnets~\cite{nath2008,kini2006}. Figure~\ref{fig:subtraction} shows the fit of $C_{\rm p}(T)$ by Eq.~\eqref{Debye} with $c_1 = 10$, $c_2 = 1$, and $c_3 = 2$, where $c_1$ is the total number of light atoms (Li and O), $c_2$ corresponds to one Cu atom per formula unit, and $c_3$ is for two W atoms per formula unit. The sum of $c_n$ is 13, which is the total number of atoms per formula unit.  Owing to the large differences in the atomic masses, we used three different Debye temperatures: $\theta_{D1}$ for Li$^{+}$ and O$^{2-}$, $\theta_{D2}$ for Cu$^{2+}$, and $\theta_{D3}$ for W$^{6+}$. One expects that the Debye temperature varies inversely with the atomic mass. Indeed, we obtained $\theta_{D1}\simeq$ 850\,K, $\theta_{D2}\simeq$ 340\,K, and $\theta_{D3}\simeq$ 200\,K. Finally, the high-$T$ fit was extrapolated down to 2.1\,K and the magnetic part $C_{\rm {mag}}(T)$ was estimated by subtracting $C_{\rm {ph}}(T)$ from $C_{\rm {p}}(T)$, see Fig.~\ref{fig:subtraction}.

In order to check the reliability of the fitting procedure, we calculated the total magnetic entropy ($S_{\rm mag}$) by integrating $C_{\rm mag}(T)/T$:
\begin{equation}
\label{smag}
S_{\rm{mag}}(T) = \int_{2.1\,\text{K}}^{T}\frac{C_{\rm{mag}}(T')}{T'}dT'.
\end{equation}
The resulting magnetic entropy is $S_{\rm{mag}}\simeq$ 5.2\,J\,mol$^{-1}$\,K$^{-1}$ at 30\,K in reasonable agreement with the expected value of $R\ln(2S+1)\simeq 5.76$\,J\,mol$^{-1}$\,K$^{-1}$ for the spin-$\frac12$ Cu$^{2+}$ ions in Li$_2$CuW$_2$O$_8$.

\begin{figure}
\includegraphics{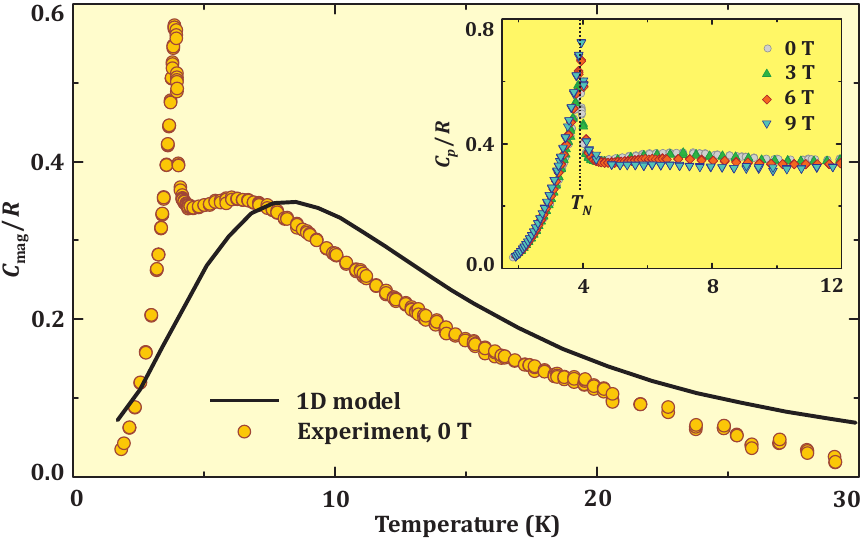}
\caption{\label{fig:heat}
(Color online) Magnetic specific heat $C_{\rm mag}$ of Li$_2$CuW$_2$O$_8$ in zero field and theoretical curve for the spin-chain model with $J_{\rm 1D}$ from the susceptibility fit. The inset shows specific heat ($C_p$) measured in different applied fields. The dotted line shows $T_N\simeq 3.9$\,K, which does not depend on the field up to 9\,T.
}
\end{figure}
Magnetic specific heat ($C_{\rm mag}$) of Li$_2$CuW$_2$O$_8$ features a broad maximum at 6\,K with the maximum value of $C_{\mg}^{\max}/R\simeq 0.35$ (Fig.~\ref{fig:heat}). At first glance, this would be again indicative of a quasi-1D scenario, but the \mbox{spin-$\frac12$} chain with $J_{\rm 1D}$ determined from the susceptibility fit should feature the maximum of $C_{\rm mag}$ at a much higher temperature. The value at the maximum, $C^{\max}$, is very sensitive to the effects of dimensionality and frustration~\cite{bernu2001}. Our data are in between those for the frustrated (triangular lattice, $C^{\max}/R\simeq 0.22$) and non-frustrated (square lattice $C^{\max}/R\simeq 0.44$) cases in 2D. Given the 3D nature of Li$_2$CuW$_2$O$_8$, this implies strong frustration that triggers quantum fluctuations, comparable to those in the 1D case of a spin chain ($C^{\max}/R\simeq 0.35$).

The $\lambda$-type anomaly in the specific heat confirms the magnetic ordering transition at $T_N\simeq 3.9$\,K in zero field. Remarkably, the ordering temperature does not depend on the magnetic field and remains at $3.90\pm 0.05$\,K for fields up to 9\,T. This contrasts with the typical behavior of both quasi-1D~\cite{moeller2009,*pan2014} and quasi-2D antiferromagnets~\cite{sengupta2009,*tsirlin2011}, where $T_N$ reveals non-monotonic field dependence. Weak fields suppress fluctuations related to the low-dimensionality, thus facilitating the formation of the long-range order, whereas stronger fields suppress the antiferromagnetic long-range order itself. The fact that in Li$_2$CuW$_2$O$_8$ we do not observe any change in $T_N$ up to 9\,T is another observation, which would not be consistent with the low-dimensional magnetic behavior, thus corroborating the 3D nature of our system.

It is worth noting that the positions of the susceptibility and specific heat maxima are mutually consistent with the quasi-1D scenario. Taking $T_{\max}$ from the $\chi$ and $C_{\mg}$ data and using theoretical results for the uniform \mbox{spin-$\frac12$} chain ($T_{\max}^{\chi}/J_{\rm 1D}\simeq 0.64$ and $T_{\max}^{C_{\mg}}/J_{\rm 1D}\simeq 0.48$~\cite{johnston2000}), we consistently arrive at $J_{\rm 1D}\simeq 13$\,K. However, this description leads to $H_s\simeq 18$\,T well below the experimental value of 29\,T. Even more importantly, when the 1D description is applied in the vicinity of the susceptibility maximum, the fit at higher temperatures fails, which should not be the case in a quasi-1D antiferromagnet, where interchain couplings manifest themselves at low temperatures, typically below $T_{\max}$. Therefore, we conclude that Li$_2$CuW$_2$O$_8$ is not a genuine quasi-1D system, but a more complex 3D frustrated antiferromagnet that mimics the quasi-1D behavior in a certain temperature range. Remarkably, the dilution behavior of this material~\cite{ranjith2015} is also reminiscent of the quasi-1D scenario, albeit with an effective interchain coupling that is way too small to account for the experimental N\'eel temperature.

\begin{figure}
\includegraphics{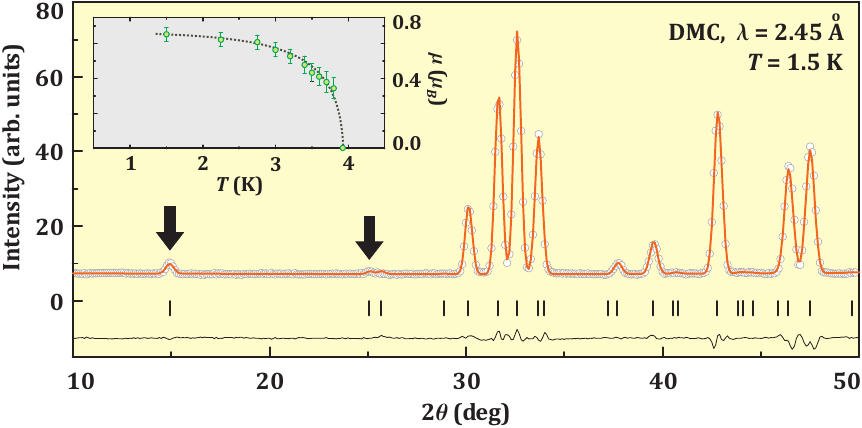}
\caption{\label{fig:neutron}
(Color online) Refined neutron diffraction pattern of Li$_2$CuW$_2$O$_8$ at 1.5\,K. The arrows denote magnetic reflections. The inset shows temperature evolution of the ordered moment ($\mu$), and the dotted line is guide for the eye.
}
\end{figure}
\textit{Magnetic ground state.} Having established that Li$_2$CuW$_2$O$_8$ is a 3D frustrated antiferromagnet, we explore its ground state by neutron diffraction. The diffraction data were collected on the DMC instrument at SINQ (PSI, Villigen) using the wavelength of 2.45\,\r A (Fig.~\ref{fig:neutron}). The nuclear scattering above $T_N$ is consistent with the triclinic room-temperature crystal structure reported in the literature~\cite{alvarez2001}. Below $T_N$, two additional magnetic reflections reveal a commensurate magnetic order with the propagation vector $\kv=(0,\frac12,0)$. For an antiferromagnet, this implies antiparallel spins along $b$ and parallel spins along $a$ and $c$ (Fig.~\ref{fig:structure}, bottom right). Magnetic moments lie in the $ac$ plane, and the size of the ordered moment is $\mu=0.65(4)$\,$\mu_B$ at 1.5\,K.

The value of the ordered moment reflects the magnitude of quantum fluctuations. In a non-frustrated 3D magnet with spin-$\frac12$, the ordered moment is about 0.83\,$\mu_B$ (cubic lattice)~\cite{schmidt2002}. The ordered moment of 0.65\,$\mu_B$ in Li$_2$CuW$_2$O$_8$ is reminiscent of a 2D case with $\mu=0.61$\,$\mu_B$~\cite{sandvik1997}, but the spin lattice of Li$_2$CuW$_2$O$_8$ has no apparent 2D features, and the reduced ordered moment should be ascribed to the effect of frustration in 3D.

The magnetic structure in the $ab$ plane is stabilized by the two stronger interactions $J_{010}$ and $J_{\bar 110}$, whereas the weaker interaction $J_{100}$ is overwhelmed. Same applies to the (effective) ferromagnetic order along the $c$ direction, where the experimental magnetic structure is stabilized by $J_{011}$ and $J_{11\bar 1}$, while the effect of the two other couplings ($J_{001}, J_{\bar 101}$) is fully suppressed. Remarkably, this order may be quantum in nature. When the spin Hamiltonian of Li$_2$CuW$_2$O$_8$:
\begin{equation}
  \hat H = \sum_{i,\rv}J_{\rv}\,\Sv_i\Sv_{i+\rv}
\end{equation}
with all seven couplings $J_{\rv}$ from Table~\ref{tab:exchanges} is considered on the classical level (i.e., with the effect of quantum fluctuations not taken into account), the ground state turns out to be incommensurate with non-collinear order along all three directions and the energy of $-8.2$\,K compared to $-7.7$\,K for the collinear state observed experimentally. This energy preference of the incommensurate state on the classical level is a robust effect that is not sensitive to uncertainties in computed exchange coupling, see Table~\ref{tab:exchanges-suppl}.

The stabilization of the collinear state by quantum fluctuations has been predicted for the spatially anisotropic spin-$\frac12$ triangular lattice~\cite{starykh2007,pardini2008,bishop2009,hauke2011,ghamari2011}, and a similar mechanism may be operative in our case. Alternatively, collinear state may be stabilized by anisotropy terms in the spin Hamiltonian. By virtue of the crystallographic symmetry, Dzyaloshinsky-Moriya interactions vanish in Li$_2$CuW$_2$O$_8$, and symmetric anisotropy remains the only plausible effect beyond quantum fluctuations that could explain the collinear ground state.

\textit{Conclusions.} Altogether, we explored frustrated 3D magnetism of Li$_2$CuW$_2$O$_8$. Density-functional calculations evidence magnetic three-dimensionality corroborated by the small curvature of the magnetization isotherm and the field-independent N\'eel temperature, whereas the shift of the susceptibility and specific heat maxima toward low temperatures as well as the reduced ordered moment of 0.65\,$\mu_B$ are indicative of strong magnetic frustration. Microscopically, this behavior is rationalized by the complex spin lattice built by triangular planes with frustrated interlayer couplings. The tangible effect of magnetic frustration and the isotropic nature of the magnetic exchange render Li$_2$CuW$_2$O$_8$ an excellent model system for quantitative theoretical analysis of frustrated quantum magnets in 3D.

\textit{Note added:} after the submission of our manuscript, Panneer Muthuselvam \textit{et al.}~\cite{panneer2015} reported another investigation of Li$_2$CuW$_2$O$_8$. From computational results they infer that the leading AFM coupling is along the $a$-direction. This contradicts the experimental magnetic structure reported in our work, where the magnetic order along $a$ is clearly ferromagnetic. Our data show that the spin lattice of Li$_2$CuW$_2$O$_8$ is quite complex and can not be captured by only three exchange couplings considered in Ref.~\onlinecite{panneer2015}. A follow-up study of the dilution behavior and the nuclear magnetic resonance data for Li$_2$CuW$_2$O$_8$ can be found in Ref.~\cite{ranjith2015}.

\begin{acknowledgments}
Financial support of DST India (RN and KMR), Mobilitas MTT77, PUT733, and Federal Ministry for Education and Research through the Sofja Kovalevkaya Award of Alexander von Humboldt Foundation (AAT), as well as of FP7 under grant agreement 290605 and by TRR80 of DFG (MS) is appreciated. We thank PSI for granting the DMC beamtime and acknowledge the support of the HLD at HZDR, member of EMFL. Fruitful discussions with Ioannis Rousochatzakis and Johannes Richter, and the usage of the HTE package~\cite{htse,*htsecode} are kindly acknowledged. 
\end{acknowledgments}

%

\end{document}